\documentclass[aps,floats]{revtex4}
\usepackage{amsmath,amssymb}
\usepackage{graphicx,epsfig}

\begin{document}
\bibliographystyle {plain}

\def\oppropto{\mathop{\propto}} 
\def\opsimeq{\mathop{\simeq}}
\def\opoverderline{\mathop{\overline}}
\def\operarrow{\mathop{\longrightarrow}}
\def\opsim{\mathop{\sim}}

\def\fig#1#2{\includegraphics[height=#1]{#2}}
\def\figx#1#2{\includegraphics[width=#1]{#2}}


\title{ Scaling of the largest dynamical barrier \\
 in the one-dimensional long-range Ising spin-glass } 


 \author{ C\'ecile Monthus and Thomas Garel }
  \affiliation{ Institut de Physique Th\'{e}orique, CNRS and CEA Saclay,
 91191 Gif-sur-Yvette, France}

\begin{abstract}
The long-range one-dimensional Ising spin-glass with random couplings decaying as $J(r) \propto r^{-\sigma}$ presents a spin-glass phase $T_c(\sigma)>0$ for $0 \leq \sigma<1$ (the limit $\sigma=0$ corresponds to the mean-field SK-model). We use the eigenvalue method introduced in our previous work [C. Monthus and T. Garel, J. Stat. Mech. P12017 (2009)] to measure the equilibrium time $t_{eq}(N)$ at temperature $T=T_c(\sigma)/2$ as a function of the number $N$ of spins. We find the activated scaling $\overline{ \ln t_{eq}(N)} \sim N^{\psi}$ with the same barrier exponent $\psi \simeq 0.33$ in the whole region $0 \leq \sigma <1$.

\end{abstract}

\maketitle

\section{ Introduction }

The dynamical properties of spin-glasses have been much studied 
since many decades (see \cite{youngbook,houches} and references therein).
In particular, the scaling of dynamical barriers in the spin-glass phase
 has been analyzed along two different paths, as we now recall. 

\subsection{ Dynamics from an initial random configuration  }

For the dynamics starting from an initial random configuration at $t=0$, 
it is useful to introduce some growing coherence length $L(t)$
with the following meaning : the smaller lengths $l < L(t)$ are quasi-equilibrated
whereas the bigger lengths $l > L(t)$  are still completely
out of equilibrium. In particular, this picture is well established for pure ferromagnets \cite{bray_revue}, where the coherence length $L(t)$ grows as the power-law
 \begin{eqnarray}
 L_{pure}(t)  && \sim t^{\frac{1}{z}}
\label{pure}
\end{eqnarray}
where $z$ is the dynamical exponent, with the numerical
value $z=2$ for non-conserved dynamics.

For disordered systems, the time-behavior of the coherence length $L(t)$
has remained controversial over the years, between a power-law similar to 
the pure case of Eq. \ref{pure}, but with a non-universal dynamical exponent $z(T,\Delta)$ that depends on the temperature $T$ and on the disorder strength $\Delta$
 \begin{eqnarray}
t  && \sim [ L(t)]^{z(T,\Delta)}
\label{nonactivated}
\end{eqnarray}
and an activated scaling with a universal 
barrier exponent $\psi^{linear}$, which does
not depend on the temperature and on the disorder strength, and characterizes the
zero-temperature fixed point of the droplet scaling theory \cite{heidelberg,Fis_Hus}
\begin{eqnarray}
\ln t  \sim [ L(t)]^{\psi^{linear}}
\label{activated}
\end{eqnarray}

For Directed Polymers in random media, the power-law behavior of Eq. \ref{nonactivated}
 has been used by various authors \cite{DP_alge}, but more recent works \cite{DP_activ}
have found the activated scaling of Eq. \ref{activated}.
For random ferromagnets, the debate between the power-law behavior \cite{randomferro_alge} and the activated scaling  \cite{randomferro_activ,nemnes} is still going on. Finally for spin-glasses, the same controversy exists 
between power-law dynamics \cite{SG_alge} 
and logarithmic dynamics \cite{SG_activ,nemnes}, as well as in the interpretation
of experimental data \cite{SG_expe}.

\subsection{ Scaling of the largest dynamical barrier  }

It turns out that another interesting dynamical observable is not controversial,
and displays activated scaling : it is the largest relaxation time $t_{eq}(N)$ needed for a finite system containing $N$ spins to converge towards thermal equilibrium.
It is also often called the ergodic time, since it represents the time needed to visit
the whole phase space of configurations. 

For instance for the pure ferromagnet in dimension $d$, 
the largest relaxation time $t^{pure}_{eq}(N)$ for a finite system containing $N=L^d$ spins corresponds near zero temperature to the time needed to go
from one ground state (where all spins take the value $+1$)
to the opposite ground state (where all spins take the value $-1$).
As a consequence, it satisfies the activated scaling
\begin{eqnarray}
\ln t^{pure}_{eq}(N=L^d)  \sim L^{d-1}= N^{\frac{d-1}{d}}
\label{ferropur}
\end{eqnarray}
where $L^{d-1}$ corresponds to the energy cost of a system-size domain-wall of dimension $(d-1)$
with respect to the ground state energy. The difference with the power-law scaling of Eq. \ref{pure} can be explained as follows.
The random initial condition is a configuration of very high energy containing
a large density of domain-walls, so that the relaxational dynamics is dominated by the
diffusion (hence the diffusion value $z=2$ for the dynamical exponent) and annihilation of the domain-walls. On the contrary, for the equilibrium time of Eq. \ref{ferropur},
one may consider that the initial condition is one of the ground state, and one is interested into the time needed to reach the opposite ground state by thermal activation~: so here the system has to {\it create a system-size domain-wall } as an intermediate step.

For the mean-field Sherrington-Kirkpatrick spin-glass model \cite{SKmodel},
which is the basis of the Replica-Symmetry-Breaking scenario \cite{replica},
there exists a consensus both theoretically \cite{rodgers,horner}
and numerically 
\cite{young,vertechi,colbourne,billoire,janke,billoire2010,billoire2011,us_conjugate}
that the largest relaxation time follows the activated scaling
\begin{eqnarray}
 \ln t^{SK}_{eq}(N)  \oppropto_{N \to \infty} N^{\psi^{SK}}
\label{activatedSK}
\end{eqnarray}
with the barrier exponent
\begin{eqnarray}
 \psi^{SK} = \frac{1}{3}
\label{psiSK}
\end{eqnarray}
Again, this largest relaxation time corresponds near zero temperature
 to the time needed to go from one ground-state to the opposite ground state.
But the scaling of Eq. \ref{activatedSK} is expected to be valid in
the whole spin-glass phase $T<T_c$.

The droplet scaling theory \cite{heidelberg,Fis_Hus} also predicts that the largest relaxation time $t_{eq}$ of a short-range spin-glass model containing $N=L^d$ spins
satisfies an activated scaling
\begin{eqnarray}
 \ln t_{eq}(N=L^d)  \oppropto_{N \to \infty} L^{\psi^{linear} } =N^{\psi}
\label{defpsi}
\end{eqnarray}
where the exponent $\psi^{linear}$ defined with respect to the length scale
is expected to coincide with the exponent of Eq. \ref{activated} as a consequence of scaling.
In the following, we will use the exponent 
$\psi=\psi^{linear}/d$ defined with respect to the total number $N$ of spins,
in order to compare more directly with the mean-field SK-model of Eq. \ref{activatedSK}
 where the notion of length does not exist.

\subsection{ Organization of the paper  }

The aim of the present paper is to measure the scaling of the largest dynamical barrier
$\ln t_{eq} (N)$ for the one-dimensional long-range Ising spin-glass \cite{kotliar} 
as a function of the number $N$ of spins, via the method
introduced in our previous work \cite{us_conjugate}.

The paper is organized as follows.
In section \ref{sec_static}, we recall the definition of the one-dimensional long-range spin-glass and its main static properties.
In section \ref{sec_dyn}, we present our numerical results concerning the dynamics.
Our conclusions are summarized in section
\ref{sec_conclusion}.

\section{ Reminder on the statics of the one-dimensional long-range spin-glass}

\label{sec_static}

Since spin-glasses on hypercubic lattices are difficult to study numerically as a function of the dimension $d$, many recent works have been devoted to one-dimensional long-range Ising Spin-Glass \cite{kotliar,KY,KYgeom,KKLH,KKLJH,Katz,KYalmeida,Yalmeida,KDYalmeida,LRmoore,KHY,KH,mori,wittmann,us_overlaptyp,us_chaos}
(note that a diluted version of the model \cite{diluted,diluted2} also exists).

\subsection { Definition of the Model}

The one-dimensional long-range Ising Spin-Glass \cite{kotliar} is defined by the
 following energy of configurations ${\cal C}$
\begin{eqnarray}
 U({\cal C}) && = - \sum_{1 \leq i <j \leq N} J_{ij} S_i S_j
\label{HSGring}
\end{eqnarray}
where the $N$ classical spins $S_i=\pm$ are placed periodically on a ring, so that the distance $r_{ij} $ between the spins $S_i$ and $S_j$ reads
\begin{eqnarray}
r_{ij}= \frac{N}{\pi} \sin \left(\vert j-i \vert \frac{\pi}{N} \right)
\label{rij}
\end{eqnarray}
The couplings are chosen to decay with respect to this distance 
as a power-law of exponent $\sigma$
\begin{eqnarray}
J_{ij}= c_N(\sigma) \frac{\epsilon_{ij}}{r_{ij}^{\sigma}}
\label{defjij}
\end{eqnarray}
where $\epsilon_{ij}$ are random Gaussian variables
of zero mean $\overline{\epsilon}=0$ and unit variance $\overline{\epsilon^2}=1 $.
The constant $c_N(\sigma) $ is defined by the condition
\begin{eqnarray}
1= \sum_{j \ne 1} \overline{J_{1j}^2} =  c_N^2(\sigma) \sum_{j \ne 1} \frac{1}{r_{1j}^{2 \sigma}}
\label{defcsigma}
\end{eqnarray}
The exponent $\sigma$ is thus the important parameter of the model.

An important critical exponent that characterizes the spin-glass phase
is the stiffness exponent $\theta$ 
associated with the difference of the ground state energies
between Periodic and Antiperiodic
boundary conditions in a given sample
(see \cite{KY} for the precise meaning of Antiperiodic
boundary conditions for the long-ranged model of Eq. \ref{HSGring})
\begin{eqnarray}
 E_{GS}^{(P)}(N)-E_{GS}^{(AP)}(N) \sim  N^{\theta} u
\label{dwegs}
\end{eqnarray}
where $u$ is a sample random variable of order $O(1)$ and of zero mean.

\subsection{Non-extensive region  $0 \leq \sigma < 1/2$ } 

\label{nonextregion}

In the non-extensive region $0 \leq \sigma < 1/2$, 
Eq. \ref{defcsigma} yields 
\begin{eqnarray}
 c_N(\sigma) \propto N^{\sigma- \frac{1}{2}}
\label{rescalcsigma}
\end{eqnarray}
so there is an explicit size-rescaling of the couplings
as in the Sherrington-Kirkpatrick (SK) 
mean-field model that corresponds to the case $\sigma=0$.
Recent studies \cite{mori,wittmann} have proposed that
both universal properties like critical exponents, but also
non-universal properties like the critical temperature 
do not depend on $\sigma$ in the whole region $0 \leq \sigma < 1/2$,
and thus coincide with the properties of the SK model $\sigma=0$.
 In particular, the exponent governing the correction to extensivity
of the averaged ground state energy 
\begin{eqnarray}
\overline{ E_{GS}^{(P)}(N)} \simeq  N e_0 +  N^{\theta_{shift}} e_1+... 
\label{defthetashift}
\end{eqnarray}
is then expected to keep the value measured in the SK model
\cite{andreanov,Bou_Krz_Mar,pala_gs,aspelmeier_MY,Katz_gs,Katz_guiding,aspelmeier_BMM,boettcher_gs,us_tails,us_matching}
\begin{eqnarray}
\theta_{shift}(0 \leq \sigma < 1/2) =\theta_{shift}^{SK} \simeq 0.33
\label{thetanonext}
\end{eqnarray}
The stiffness exponent $\theta(\sigma)$ of Eq. \ref{dwegs}
measured in \cite{KY} are compatible with this constant value.

The critical temperature is also expected to remain constant 
in the whole non-extensive region $0 \leq \sigma < 1/2$ \cite{mori,wittmann}
\begin{eqnarray}
T_c(0 \leq \sigma < 1/2) = T_c^{SK} =1
\label{tcnonext}
\end{eqnarray}

\subsection{Extensive region $\sigma>1/2$  } 

\label{extregion}

In the extensive region $\sigma>1/2$, Eq. \ref{defcsigma} yields 
\begin{eqnarray}
 c_N(\sigma) =O(1)
\label{norescalcsigma}
\end{eqnarray}
so there is no size rescaling of the couplings.
The limit $\sigma =+\infty$ corresponds to the short-range one-dimensional model.
There exists a spin-glass phase at low temperature for $\sigma<1$ \cite{kotliar}.
The stiffness exponent $\theta(\sigma)$  of Eq. \ref{dwegs}
measured via Monte-Carlo simulations
on sizes $16 \leq N \leq 256$ in \cite{KY} decays as $\sigma$ grows
with the following values (see \cite{KY} for other values of $\sigma$) 
\begin{eqnarray}
\theta(\sigma=0.62) && \simeq 0.24
\nonumber \\
\theta(\sigma=0.75) && \simeq 0.17
\nonumber \\
\theta(\sigma=0.87) && \simeq 0.08
\nonumber \\
\theta(\sigma=1) && \simeq 0
\label{thetaext}
\end{eqnarray}
These values
can be recovered on smaller sizes $6 \leq L \leq 24$ via exact enumeration 
\cite{us_overlaptyp}, by considering the change between Periodic and Antiperiodic Boundary Conditions, or by considering the correction to extensivity of the averaged ground state energy.

The critical point  is mean-field-like for $\sigma<2/3$, and non-mean-field-like for $2/3<\sigma<1$ \cite{kotliar}. The critical temperature decays as $\sigma$ grows, with the following values
\cite{KYalmeida}
\begin{eqnarray}
T_c(\sigma=0.65) && \simeq 0.86
\nonumber \\
T_c(\sigma=0.75) && \simeq 0.69
\nonumber \\
T_c(\sigma=0.85) && \simeq 0.49
\nonumber \\
T_c(\sigma=1) && \simeq 0
\label{tcext}
\end{eqnarray}

\subsection{Comparison with the short-range spin-glass in dimension $d$   } 

The critical exponents for the spin-glass transition at $T_c$
have been compared between the long-range model (LR)
in one dimension and the short-range (SR) model in dimension $d$
 \cite{KDYalmeida,diluted2} with the following conclusion (see \cite{KDYalmeida,diluted2} for more details) : the SR model in dimension $d=3$ is somewhat 
similar to the LR model
of parameter $\sigma \simeq 0.896$, whereas the LR model in dimension $d=4$ 
is somewhat similar to the LR model of parameter $\sigma \simeq 0.79$.
The mean-field region $d \geq d_c = 6$ is similar to the mean-field region 
$\sigma < \sigma_c=\frac{2}{3}$ with $d=2/(2 \sigma-1)$.

One may also establish some correspondence based on the stiffness exponent 
of Eq. \ref{dwegs} which characterizes the zero-temperature fixed point.
For the SR model in dimension $d$, 
the values measured for the stiffness exponent
$\theta^{SR}_{linear}$ defined with respect to the linear length $L$
(see \cite{boettcher} and references therein)
reads for the stiffness exponent $\theta^{SR}=\theta^{SR}_{linear}/d$ 
of Eq \ref{dwegs} defined with respect to the number $N=L^d$ of spins
\begin{eqnarray}
\theta^{SR}(d=2) && \simeq - \frac{0.28}{2} \simeq -0.14
\nonumber \\
\theta^{SR}(d=3) && \simeq \frac{0.24}{3} \simeq 0.08
\nonumber \\
\theta^{SR}(d=4) && \simeq  \frac{0.61}{4} \simeq 0.15
\nonumber \\
\theta^{SR}(d=5) && \simeq  \frac{0.88}{5} \simeq 0.176
\nonumber \\
\theta^{SR}(d=6) && \simeq \frac{1.1}{6} \simeq 0.183
\label{hypercubic}
\end{eqnarray}
that may be compared to the values of the stiffness model
$\theta(\sigma)$ for the LR model as a function of $\sigma$
(see Eq. \ref{thetaext}).

\section{ Scaling of the largest dynamical barrier in the spin-glass phase }

\label{sec_dyn}

In this section, we explain how the largest dynamical barrier
can be obtained from the Master Equation defining the stochastic dynamics.

\subsection{ Metropolis dynamics }

The Metropolis dynamics of the long-range spin-glass of Eq. \ref{HSGring}
is defined by the Master Equation
\begin{eqnarray}
\frac{ dP_t \left({\cal C} \right) }{dt}
= \sum_{\cal C '} P_t \left({\cal C}' \right) 
W \left({\cal C}' \to  {\cal C}  \right) 
 -  P_t \left({\cal C} \right) \sum_{ {\cal C} '} W \left({\cal C} \to  {\cal C}' \right) 
\label{master}
\end{eqnarray}
where the transition rate $ W \left({\cal C}' \to  {\cal C}  \right) $  from configuration 
${\cal C}'$ to ${\cal C}$ 
reads in terms of the energy of Eq. \ref{HSGring}
\begin{eqnarray}
W \left( \cal C \to \cal C '  \right)
= \delta_{<\cal C, \cal C' >} 
{\rm min} \left[1, e^{-  \frac{(U({\cal C' })-U({\cal C }))}{T}} \right]
\label{metropolis}
\end{eqnarray}
The notation $\delta_{<\cal C, \cal C' >} $ means that the two configurations are
related by a single spin-flip.
These rates satisfy the detailed-balance property
\begin{eqnarray}
 P_{eq}({\cal C})  W \left( \cal C \to \cal C '  \right)
= P_{eq}({\cal C'})  W \left( \cal C' \to \cal C   \right)
\label{detailed}
\end{eqnarray}
with respect to the Boltzmann thermal equilibrium 
\begin{eqnarray}
P_{eq}({\cal C}) && = \frac{ e^{- \frac{U({\cal C})}{T}} }{Z }
\nonumber \\
Z &&  = \sum_{\cal C}  e^{- \frac{U({\cal C})}{T}} 
\label{equi}
\end{eqnarray}

\subsection{ Associated quantum Hamiltonian  }

As is well known (see for instance the textbooks \cite{gardiner,vankampen,risken})
 any master equation satisfying detailed-balance
 can be transformed into a symmetric operator
 via the change of variable
\begin{eqnarray}
P_t ( {\cal C} ) \equiv    e^{-  \frac{U(\cal C )}{2T} } \psi_t  ( {\cal C} )
\label{relationPpsi}
\end{eqnarray}
The function $\psi_t  ( {\cal C} )$ then satisfies the imaginary-time  Schr\"odinger
equation
\begin{eqnarray}
\frac{ d\psi_t \left({\cal C} \right) }{dt} = -H \psi_t \left({\cal C} \right)
\label{Hquantum}
\end{eqnarray}
where the quantum Hamiltonian reads in configuration space
\begin{eqnarray}
 H = \sum_{\cal C } \epsilon \left( {\cal C} \right) \vert {\cal C} > < {\cal C } \vert
+ \sum_{{\cal C},{\cal C '}}  V({\cal C} , {\cal C' })
 \vert {\cal C} > < {\cal C' } \vert
\label{tight}
\end{eqnarray}
The on-site energies read
\begin{eqnarray}
 \epsilon \left( {\cal C} \right) = \sum_{ {\cal C} '} W \left({\cal C} \to  {\cal C}' \right)
\label{eps}
\end{eqnarray}
whereas the hopping terms read
\begin{eqnarray}
 V({\cal C} , {\cal C' })=- e^{-  \frac{(U(\cal C' )-U(\cal C ))}{2T} } W \left( \cal C' \to \cal C   \right)
\label{hopping}
\end{eqnarray}

For spin models without disorder,
 this mapping has been used for more than fifty years
\cite{glauber,felderhof,siggia,kimball,peschel,Night_Blo,us_dynquantumrg,us_dyndyson}.
For disordered systems, this mapping onto a quantum hamiltonian 
has been also much used both for the one-dimensional diffusion in random media \cite{jpbreview,laloux,golosovloc,texier} and for spin models like the Sherrington-Kirkpatrik model
(\cite{us_conjugate} and Appendix B of \cite{castelnovo}) or disordered ferromagnets
\cite{us_dyncayleyrandom,us_dynferroMK}.

\subsection{ Properties of the spectrum of the quantum Hamiltonian $H$ }

The spectral decomposition of the evolution operator $e^{-t H}$
associated to the quantum Hamiltonian of Eq. \ref{tight}
reads in terms of the eigenvalues $E_n$ and the associated normalized eigenvectors $\vert \psi_n>$
\begin{eqnarray}
  e^{-t H}  && = \sum_{n \geq 0} e^{-t E_n } \vert \psi_n>< \psi_n \vert 
\label{spectreH}
\end{eqnarray}
So the conditional probability 
$P_t \left( {\cal C} \vert {\cal C}_0\right)$ to be in configuration ${\cal C}$ at $t$
if one starts from the configuration ${\cal C}_0$ at time $t=0$ can be written as
\begin{eqnarray}
P_t \left( {\cal C} \vert {\cal C}_0\right) =
   e^{-  \frac{U({\cal C} )-U({\cal C}_0 )}{2T} } <{\cal C} \vert e^{-t H}  \vert {\cal C}_0>
= e^{-  \frac{U({\cal C} )-U({\cal C}_0 )}{2T} }
\sum_n e^{- E_n t} \psi_n({\cal C})\psi_n^*({\cal C}_0)
\label{expansionP}
\end{eqnarray}

The quantum Hamiltonian $H$ has actually very special properties
which come from its relation to the dynamical master equation :

(i) the ground state energy vanishes $E_0=0$, and the corresponding
eigenvector is exactly known to be
\begin{eqnarray}
\psi_0 ({\cal C}) = \frac{ e^{- \frac{U({\cal C})}{2T}} }{\sqrt Z}
\label{psi0}
\end{eqnarray}
where $Z$ is the partition function of Eq. \ref{equi}.

This corresponds to the convergence towards the Boltzmann equilibrium in
 Eq. \ref{relationPpsi} for any initial condition ${\cal C}_0$
\begin{eqnarray}
P_t \left( {\cal C} \vert {\cal C}_0\right)
\opsimeq_{t \to + \infty}  e^{-  \frac{U({\cal C} )-U({\cal C}_0 )}{2T} }
\psi_0({\cal C})\psi_0^*({\cal C}_0) = \frac{e^{- \frac{U({\cal C})}{T}}}{Z} = P_{eq}({\cal C})
\label{CVeqP}
\end{eqnarray}

(ii) the other energies $E_n>0$ determine the relaxation towards equilibrium.
In particular, the lowest non-vanishing energy $E_1$
determines the largest relaxation time $(1/E_1)$ of the system 
\begin{eqnarray}
P_t \left( {\cal C} \vert {\cal C}_0\right) - P_{eq}({\cal C})
\opsimeq_{t \to + \infty} e^{- E_1 t}  e^{-  \frac{U({\cal C} )-U({\cal C}_0 )}{2T} }
\psi_1({\cal C})\psi_1^*({\cal C}_0) 
\label{CVeqP1}
\end{eqnarray}
This largest relaxation time represents the 'equilibrium time'
discussed in the introduction
\begin{eqnarray}
t_{eq} \equiv \frac{1}{E_1}
\label{deftaueq}
\end{eqnarray}

In summary, the largest relaxation time $t_{eq}$ 
can be computed without simulating the dynamics
by any eigenvalue method able
to compute the first excited energy $E_1$ of the quantum Hamiltonian $H$
(where the ground state is given by Eq. \ref{psi0} and has for eigenvalue $E_0=0$).

\subsection{ Conjugate gradient method in each sample to compute $E_1$ }

\label{secconjugate}

The 'conjugate gradient method' has been introduced as an iterative algorithm 
to find the minimum of functions of several variables with much better convergence
properties than the 'steepest descent' method \cite{shewchuk,golub}.
It can be applied to find the ground state eigenvalue and the associated eigenvector
by minimizing the corresponding Rayleigh quotient \cite{bradbury,nightingaleCG}
\begin{eqnarray}
R \equiv \frac{<v \vert H \vert v>}{<v \vert  v>}
\label{rayleigh}
\end{eqnarray}
In our previous work \cite{us_conjugate}, we have proposed to adapt the method described in 
\cite{bradbury,nightingaleCG} concerning the ground state $E_0$
to compute instead the first excited energy $E_1$ : the only change is that
the  Rayleigh quotient has to be minimized
 within the space orthogonal to the ground state.
More precisely, in spin models where there is a global symmetry under a global flip of all the spins, the ground state $\psi_0$
of Eq. \ref{psi0} is symmetric under a global flip of all the spins,
whereas the first excited state $\psi_1$ is anti-symmetric 
under a global flip of all the spins.
As a consequence, it is convenient to choose
the initial trial eigenvector $\vert v >$ for the conjugate gradient method
 as follows : 
denoting ${\cal C}_{pref}=\{ S_i^{pref} \}$ and ${\widehat {\cal C}}_{pref}=\{ - S_i^{pref} \}$ the two opposite configurations where the ground state $\psi_0$ of Eq. \ref{psi0} is maximal,
one introduces the overlap between an arbitrary configuration ${\cal C}$ and ${\cal C}_{pref}$
\begin{eqnarray}
Q({\cal C},{\cal C}_{pref}) =  \sum_{i=1}^N S_i S_i^{pref}
\label{overlap}
\end{eqnarray}
and the vector
\begin{eqnarray}
v({\cal C}) = {\rm sgn } \left(Q({\cal C},{\cal C}_{pref}) \right) \psi_0( {\cal C})
\label{vinitialconjugate}
\end{eqnarray}
This vector is anti-symmetric under a global flip of all the spins and thus orthogonal
to the ground state $\psi_0$. Moreover, it has already a small Rayleigh quotient
(Eq. \ref{rayleigh}) because within each valley where the sign of the overlap is fixed,
it coincides up to a global sign with the ground state $\psi_0$ of zero energy.
So the non-zero value of the Rayleigh quotient of Eq. \ref{rayleigh} only comes from 
configurations of nearly zero overlap $Q$. As a consequence,
 it is a good starting point
for the conjugate gradient method to converge rapidly towards
 the first excited state $\psi_1$.

We have applied this procedure to the long-range spin-glass model of Eq. \ref{HSGring}
for various values of $\sigma$. We have chosen to work in the 'middle' of the spin-glass phase, i.e.  at the temperature 
\begin{eqnarray}
T=\frac{T_c(\sigma)}{2}
\label{thalf}
\end{eqnarray}
where the values of the critical temperature $T_c(\sigma)$ 
obtained in Ref \cite{KYalmeida} have been recalled in Eq. \ref{tcext}.
The data presented below have been obtained for systems
of $6 \leq N \leq 20$ spins (the space of configurations is of size $2^N$), with 
the following statistics of $n_s(N)$ independent disordered samples
\begin{eqnarray}
N && = 6; 8; 10,12;14;16;18; 20 \nonumber \\
n_s(N) && = 2.10^8;2.10^8; 36.10^6 ; 63.10^5;96.10^4; 16.10^4; 28.10^3;3.10^3
\label{nume}
\end{eqnarray}
These sizes are small with respect to the sizes $N \leq 1024 $ used in Monte-Carlo
simulations to measure the largest barrier of the SK model
\cite{billoire,janke,billoire2010,billoire2011}, but we have checked in 
our previous work \cite{us_conjugate} that 
the sizes and statistics of Eq. \ref{nume} allow to recover
the correct exponent $ \psi^{SK} = \frac{1}{3}$ measured in 
Monte Carlo simulations on larger sizes.
The large statistics of Eq. \ref{nume} also
allows to better characterize the histogram
over samples.

The output is the histogram $Q_N(\Gamma_{eq})$ over disordered samples of the
largest dynamical barrier (Eq. \ref{deftaueq})
\begin{eqnarray}
\Gamma_{eq} \equiv \ln t_{eq} = - \ln E_1
\label{defbarrierflip}
\end{eqnarray}

The simplest observable is the averaged value, that defines the barrier exponent $\psi$ of Eq. \ref{defpsi}
\begin{eqnarray}
\overline{\Gamma_{eq}(N)} \equiv \overline{ \ln t_{eq}(N) } \oppropto_{N \to \infty} N^{\psi}
\label{barrierav}
\end{eqnarray}
but it is of course also interesting to consider the width $\Delta(N)$ of the probability distribution $Q_N(\Gamma_{eq})$, 
that defines the sample-to-sample fluctuation exponent $\psi_{width}$
\begin{eqnarray}
\Delta(N) \equiv
 \left( \overline{\Gamma_{eq}^2}(N)
 - (\overline{\Gamma_{eq}}(N))^2\right)^{1/2} 
  \oppropto_{N \to \infty} N^{\psi_{width}}
\label{barrierwidth}
\end{eqnarray}
For the SK model, the introduction of a different exponent for the width 
$\psi_{width} \simeq 0.25<\psi=1/3$ has been proposed in Ref. \cite{janke},
but another work \cite{billoire2010} is in favor of the same exponent 
$\psi_{width} \simeq 0.33 \simeq \psi$.

Finally, we will consider the rescaled distribution ${\tilde Q}$
of the reduced variable $u \equiv \frac{\Gamma_{eq} - \overline{\Gamma_{eq}}(N) }{\Delta(N) }$ 
\begin{eqnarray}
Q_{N}(\Gamma_{eq}) \sim  
  \frac{1}{\Delta(N) } {\tilde Q} 
\left( u \equiv \frac{\Gamma_{eq} - \overline{\Gamma_{eq}}(N) }{\Delta(N) }
 \right)
\label{defQrescaled}
\end{eqnarray}
and in its asymptotic behavior
\begin{eqnarray}
\ln {\tilde Q} (u) \oppropto_{u \to + \infty} - u^{\eta}
\label{defeta}
\end{eqnarray}
that defines the tail exponent $\eta$.

\begin{figure}[htbp]
 \includegraphics[height=6cm]{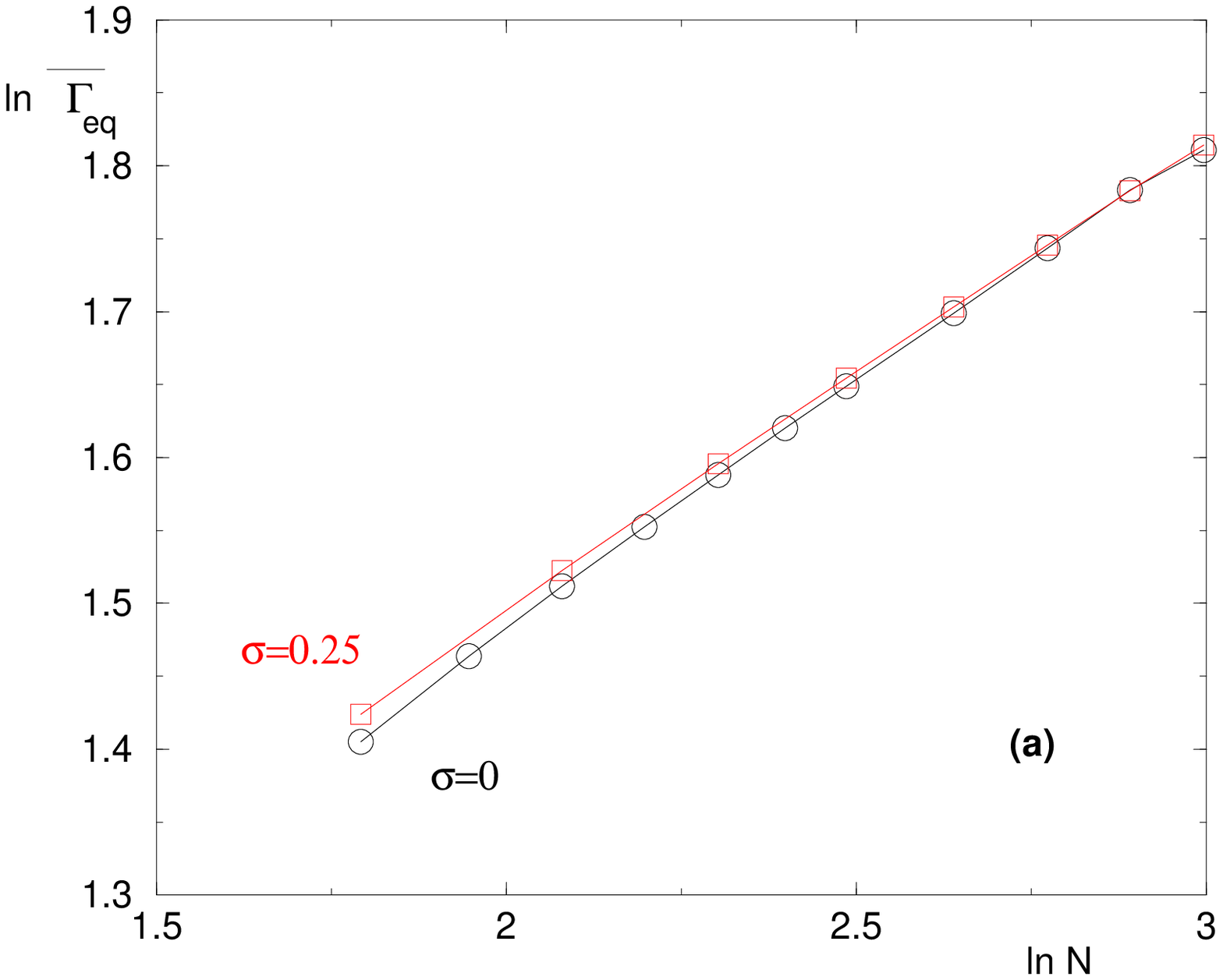}
\hspace{2cm}
 \includegraphics[height=6cm]{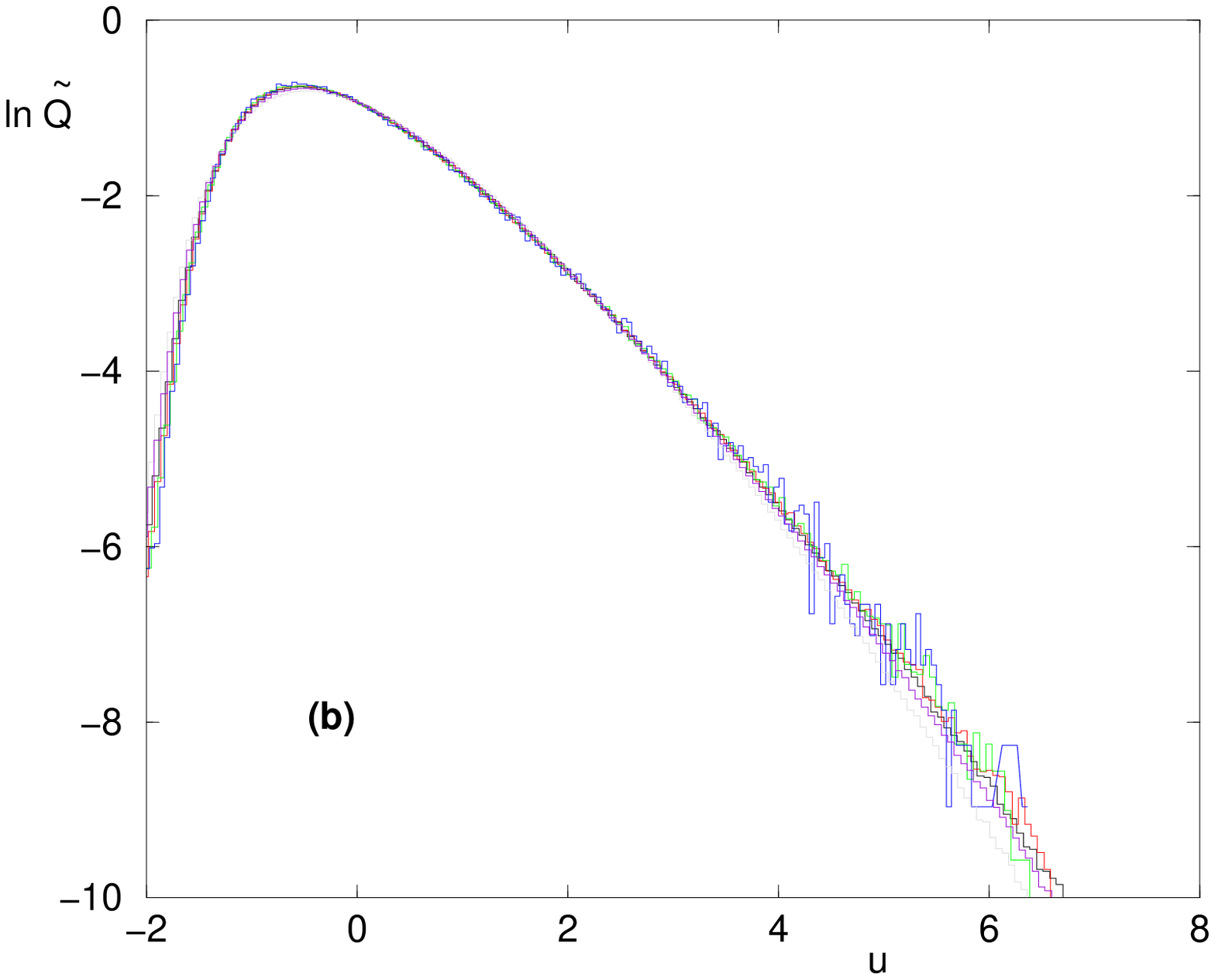}
\vspace{1cm}
\caption{ Statistics over samples of the largest barrier $\Gamma_{eq}(N) \equiv
 \ln t_{eq}(N) $ for the long-range one-dimensional spin-glass of $N$ spins
with $\sigma=0.25$ (middle of the non-extensive region $0 \leq \sigma < 1/2$)
(a) Scaling of the averaged value $\overline{\Gamma_{eq}(N)} \propto N^{\psi} $ 
in a log-log plot : the slope corresponds to the
 barrier exponent $\psi \simeq 0.33$. Note that the data 
 for $\sigma=0.25$ almost coincide with the data concerning the SK model ($\sigma=0$)
(b) Rescaled probability distribution  ${\tilde Q}(u )$ of Eq. \ref{defQrescaled}
for the sizes $N=6;8;10;12;14;16$
shown here in log-scale to see the tail of Eq. \ref{defeta} : 
the tail exponent is of order $\eta \simeq 1.33$. }
\label{fignonext}
\end{figure}

\subsection{ Numerical results in the non-extensive region $0 \leq \sigma < 1/2$ }

As recalled in section \ref{nonextregion}, the static properties are expected
to remain the same in the whole
non-extensive region $0 \leq \sigma < 1/2$ \cite{mori,wittmann}.
Our present numerical results for the dynamics are also compatible
 with this statement.
As shown on Fig.  \ref{fignonext} (a) concerning the average value 
$\overline{\Gamma_{eq}(N)}  $,
 the data concerning $\sigma=0.25$ in the middle 
of the non-extensive region nearly coincide (especially for the largest sizes)
with our previous data concerning the SK model \cite{us_conjugate}
corresponding to $\sigma=0$. Our conclusion is thus that the barrier exponent 
$\psi$ of Eq. \ref{barrierav}
keeps the value of Eq. \ref{psiSK} in the whole non-extensive region
\begin{eqnarray}
 \psi(0 \leq \sigma < 1/2) = \psi^{SK} \simeq 0.33
\label{psinonext}
\end{eqnarray}
Note that this value also coincides with the exponent $\theta_{shift}(0 \leq \sigma < 1/2) \simeq 0.33$ of Eq. \ref{thetanonext} and with the stiffness exponent $\theta(\sigma) \simeq 0.33$ measured in \cite{KY}
\begin{eqnarray}
 \psi( \sigma)  \simeq 0.33 \simeq \theta(\sigma) 
\label{psithetanonext}
\end{eqnarray}

As shown on Fig. \ref{fignonext} (b),  the probability distribution
$Q_N(\Gamma_{eq})$
convergences rapidly towards the fixed rescaled distribution ${\tilde Q}(u)$ 
of Eq. \ref{defQrescaled} : the corresponding tail exponent $\eta$ of Eq. \ref{defeta}
is of order
\begin{eqnarray}
 \eta(0 \leq \sigma < 1/2 )  \simeq 1.33
\label{etathetanonext}
\end{eqnarray}

\subsection{ Numerical results in the extensive region $1/2 < \sigma < 1$ }

\begin{figure}[htbp]
 \includegraphics[height=6cm]{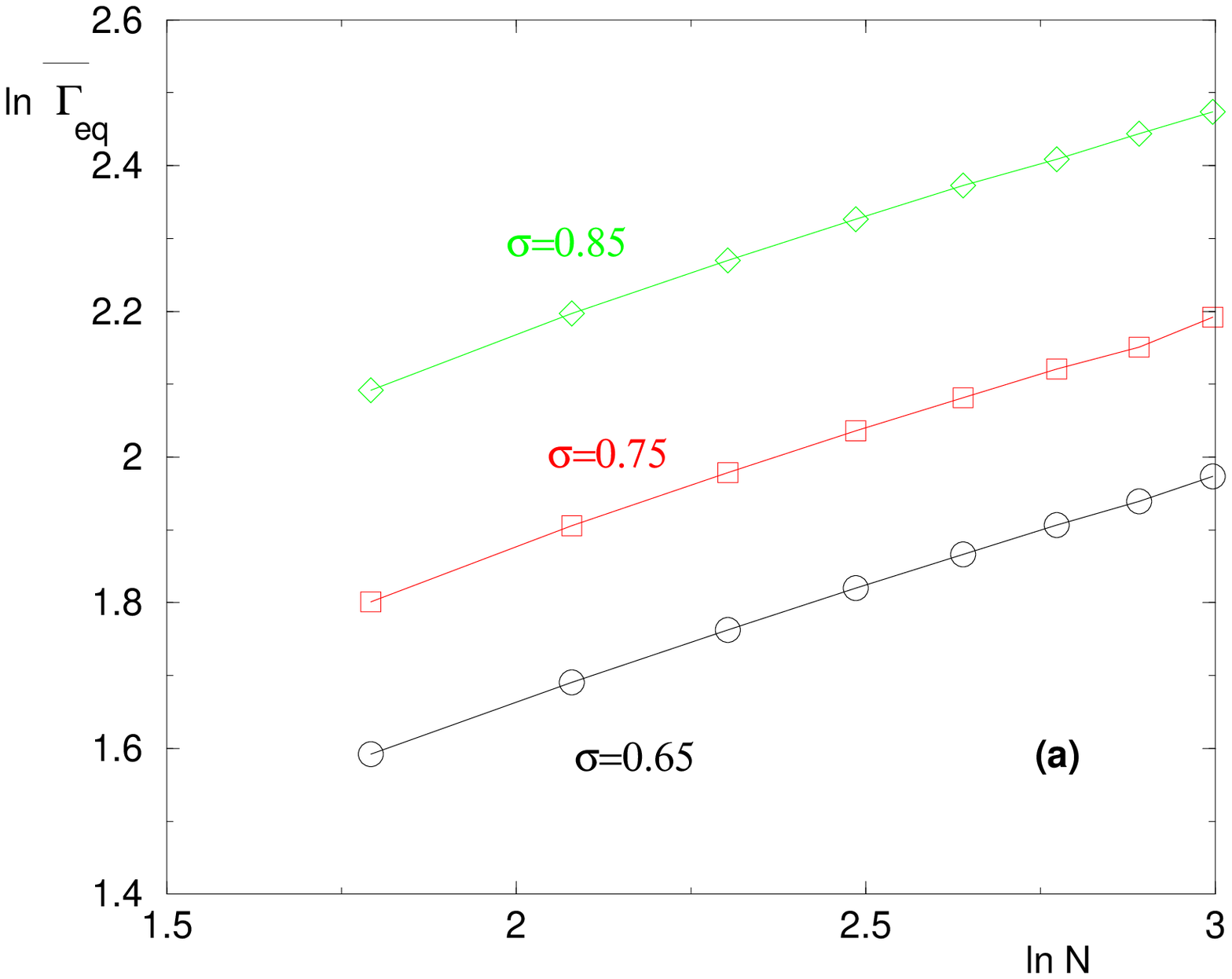}
\hspace{2cm}
 \includegraphics[height=6cm]{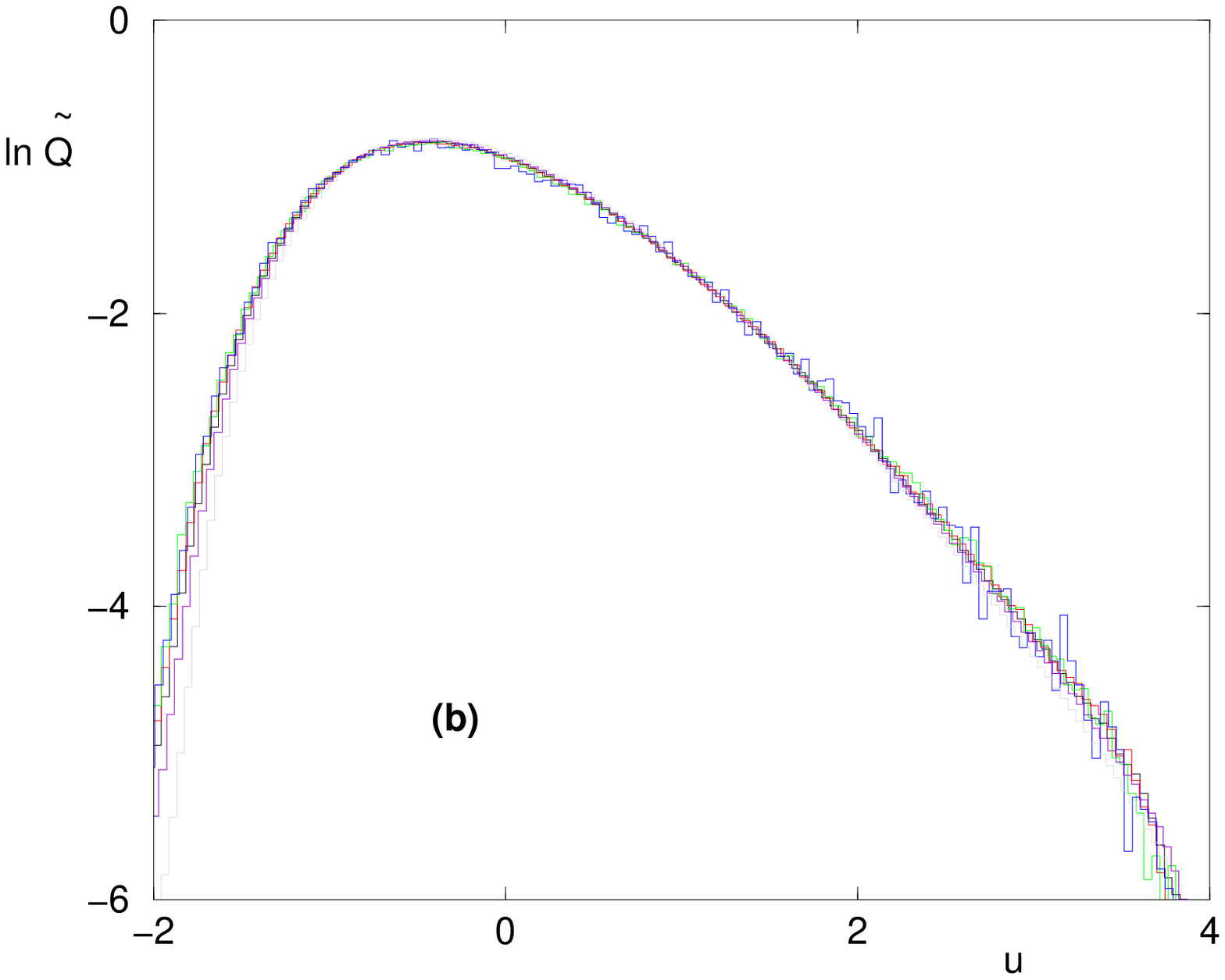}
\vspace{1cm}
\caption{ 
Statistics over samples of the largest barrier $\Gamma_{eq}(N) \equiv
 \ln t_{eq}(N) $ for the long-range one-dimensional spin-glass of $N$ spins
in the extensive region $ \sigma > 1/2$)
(a) Scaling of the averaged value $\overline{\Gamma_{eq}(N)} \propto N^{\psi} $ 
in a log-log plot for  $\sigma=0.65$, $\sigma=0.75$ and for $\sigma=0.85$ : 
the three slopes correspond to the
 same barrier exponent of order $\psi \simeq 0.33$.
(b) Rescaled probability distribution  ${\tilde Q}(u )$ of Eq. \ref{defQrescaled}
for $\sigma=0.75$ for the sizes $N=6;8;10;12;14;16$
shown here in log-scale to see the tail of Eq. \ref{defeta} : 
the tail exponent is of order $ \eta (\sigma=0.75) \simeq 1.45$. }
\label{figext}
\end{figure}

As shown on Fig. \ref{figext} (a) concerning the average value 
$\overline{\Gamma_{eq}(N)}  $, our data 
for $\sigma=0.65, 0.75, 0.85$ in the extensive region $1/2 < \sigma < 1$
 correspond to the same slope with the same value
of the barrier exponent as in Eq. \ref{psinonext}
\begin{eqnarray}
 \psi(1/2 < \sigma < 1)  \simeq 0.33
\label{psiext}
\end{eqnarray}
This result is rather surprising, since the static properties,
and in particular the stiffness exponent $\theta(\sigma)$ (Eq. \ref{thetaext}),
 decays continuously as $\sigma$ grows in the extensive region.
Using the numerical values of Eq. \ref{thetaext} obtained in Ref \cite{KY},
 we note that the inequality
\begin{eqnarray}
 \psi( \sigma )  \simeq 0.33 >  \theta(\sigma) 
\label{psithetaext}
\end{eqnarray}
is satisfied,
in agreement with the general bound $\psi \geq \theta$ of the droplet scaling theory
\cite{Fis_Hus}. 

As shown on Fig. \ref{figext} (b) for the value $\sigma=0.75$,
  the probability distribution $Q_N(\Gamma_{eq})$
again convergences rapidly towards the fixed rescaled distribution ${\tilde Q}(u)$ 
of Eq. \ref{defQrescaled}.
From our data for the three value $\sigma=0.65;0.75;0.85$,
  the tail exponent $\eta(\sigma)$ of Eq. \ref{defeta}
seems to slightly grow with $\sigma$. 
It is not clear to us whether it is only a numerical artefact or a real effect.
Nevertheless, we can at least state that it remains in the interval 
\begin{eqnarray}
1.33 \leq \eta(1/2 \leq \sigma < 1 )  \leq 1.5
\label{etathetaext}
\end{eqnarray}

\subsection{ Discussion}

In a previous work \cite{us_conjecture}, we have proposed
to interpret the difference between the dynamical exponent $\psi$ and 
the stiffness exponent $\theta$ as follows. 
The stiffness exponent $\theta$ defined by Eq. \ref{dwegs}
at zero temperature actually characterizes the whole 
finite temperature spin-glass phase,
where it governs the free-energy difference between Periodic and Antiperiodic
boundary conditions
\begin{eqnarray}
 F^{(P)}(N)-F^{(AP)}(N) \sim  N^{\theta} u
\label{dwfree}
\end{eqnarray}
As a consequence, it is the result of a global optimization over the whole system, whereas the local dynamics cannot maintain the global optimization at all times, and sees individual configurations, so that the dynamical exponent $\psi$ should actually coincide with the exponent $\theta_S$ 
\begin{eqnarray}
{\rm conjecture : }\ \ \ \ \psi = \theta_S
\label{conjecture}
\end{eqnarray}
which governs the scaling of the entropy difference 
in a given sample between Periodic and Antiperiodic boundary conditions
\begin{eqnarray}
S^{(P)}(N)-S^{(AP)}(N) \sim N^{\theta_S} v
\label{entropy}
\end{eqnarray} 
where $v$ is a sample Gaussian random variable of order $O(1)$.

In short-range models, the droplet scaling theory \cite{heidelberg,Fis_Hus}
predicts that the entropy of extensive droplets 
scales as $ L^{\frac{d_s}{2}}$ (coming from some Central Limit Theorem for independent local contributions along the interface of fractal dimension $d_s$)
\begin{eqnarray}
{\rm SR} :\ \ \  \theta_S^{linear} = \frac{d_s}{2}
\label{thetaentropylinear}
\end{eqnarray} 
The conjecture of Eq. \ref{conjecture}
\begin{eqnarray}
\psi_{SR}^{linear}=\theta_S^{linear} = \frac{d_s}{2}
\end{eqnarray}
has been discussed in detail in \cite{us_conjecture} for spin-glasses,
but also for other disordered models like directed polymers or random ferromagnets.

For the one-dimensional long-range model, the entropy exponent $\theta_S$
of Eq. \ref{entropy} has been numerically measured in our recent work \cite{us_chaos} :
the conclusion is that it
takes the same simple value both in the non-extensive region $0\leq \sigma<1/2$
 and in the extensive region $1/2 < \sigma <1$
\begin{eqnarray}
{\rm LR} :\ \ \   \theta_S(0 \leq \sigma \leq 1) \simeq \frac{1}{3}
\label{thetaS}
\end{eqnarray}
Our conclusion is thus that the conjecture of Eq. \ref{conjecture}
is satisfied with the same constant value found here for the dynamical exponent
\begin{eqnarray}
{\rm LR} : \ \ \  \psi(0 \leq \sigma \leq 1) \simeq \frac{1}{3} \simeq
\theta_S(0 \leq \sigma \leq 1)
\label{psithetaS}
\end{eqnarray}
However, there exists an important
 difference between the non-extensive region and the extensive region :

(i) in the non-extensive region $0\leq \sigma<1/2$, the 
stiffness exponent $\theta$ also takes the same value $1/3$ 
\begin{eqnarray}
0\leq \sigma<1/2 : \psi \simeq \theta_S \simeq \frac{1}{3} \simeq \theta
\label{psithetaSnonext}
\end{eqnarray}

(ii) in the non-extensive region $\sigma>1/2$, the stiffness exponent  $\theta(\sigma)$
takes smaller values (see Eq. \ref{thetaext})
\begin{eqnarray}
1/2 \leq \sigma<1 : \psi \simeq \theta_S \simeq \frac{1}{3} > \theta (\sigma)
\label{psithetaSext}
\end{eqnarray}
So here there is an entropy-energy cancellation mechanism 
with respect to the free-energy as
in short-ranged models \cite{heidelberg,Fis_Hus}.

\section{ Conclusion }

\label{sec_conclusion}

In this paper, we have used the eigenvalue method introduced in our previous work \cite{us_conjugate} to measure the scaling of the largest dynamical barrier for the 
 long-range one-dimensional Ising spin-glass as a function of the parameter $\sigma$.
In the whole region where a spin-glass phase exists, 
we have found the same barrier exponent
\begin{eqnarray}
 \psi(0 \leq \sigma < 1)  \simeq 0.33
\label{psifull}
\end{eqnarray} 
We have proposed that this value coincides with the entropy exponent $\theta_S(\sigma)$ recently measured in \cite{us_chaos}, in agreement with the general conjecture
of Eq. \ref{conjecture}. 

 If the simple value of Eq. \ref{psifull}
is confirmed in the future by other numerical studies on larger sizes (like the references \cite{billoire,janke,billoire2010,billoire2011} concerning the SK model corresponding to the special case $\sigma=0$), we hope that it will help to better understand the phase space structure of the spin-glass phase.

\end{document}